\pdfoutput=1
\documentclass[conference]{IEEEtran}
\usepackage{tabularx}
\usepackage{times}
\usepackage{epsfig}
\usepackage{graphicx}
\usepackage{amsmath}
\usepackage{amssymb}
\usepackage{arydshln}
\usepackage{wasysym}
\usepackage{algorithm}
\usepackage[noend]{algpseudocode}
\usepackage{tabto}
\usepackage{enumitem}
\usepackage{subcaption}

\begin{document}

\title{Ballistocardiogram-based Authentication using Convolutional Neural Networks}


\author{\IEEEauthorblockN{Joshua Hebert}
\IEEEauthorblockA{
Worcester Polytechnic Institute\\
Worcester, MA, 01609\\
Email: jahebert@wpi.edu}
\and
\IEEEauthorblockN{Brittany Lewis}
\IEEEauthorblockA{
Worcester Polytechnic Institute\\
Worcester, MA, 01609\\
Email: bfgradel@wpi.edu}
\and
\IEEEauthorblockN{Hang Cai}
\IEEEauthorblockA{
Worcester Polytechnic Institute\\
Worcester, MA, 01609\\
Email:hcai@wpi.edu}
\and
\IEEEauthorblockN{ Krishna K. Venkatasubramanian}
\IEEEauthorblockA{
Worcester Polytechnic Institute\\
Worcester, MA, 01609\\
Email: kven@wpi.edu}
\and
\IEEEauthorblockN{Matthew Provost}
\IEEEauthorblockA{TechAccess of Rhode Island\\
Cranston, RI, 02921 \\
Email: matthewp@techaccess-ri.org .edu}
\and
\IEEEauthorblockN{Kelly Charlebois}
\IEEEauthorblockA{TechAccess of Rhode Island\\
Cranston, RI, 02921 \\
Email: kellyc@techaccess-ri.org .edu}}

\maketitle

%%%%%%%%% ABSTRACT
\begin{abstract}
%There has been an explosion in the introduction of head-based wearable internet-of-things (WIoT) in recent years. One of the potential uses of such head-based WIoT devices is providing new sources of biometrics for authentication. 
The goal of this work is to demonstrate the use of the {\em ballistocardiogram (BCG)} signal, derived using head-mounted wearable devices, as a viable biometric for authentication. The BCG signal is the measure of an person's body acceleration as a result of the heart's ejection of blood. It is a characterization of the cardiac cycle and can be derived non-invasively from the measurement of subtle movements of a person's extremities. In this paper, we use several versions of the BCG signal, derived from accelerometer and gyroscope sensors on a Smart Eyewear (SEW) device, for authentication. The derived BCG signals are used to train a convolutional neural network (CNN) as an authentication model, which is personalized for each subject. We evaluate our authentication models using data from 12 subjects and show that our approach has an equal error rate (EER) of 3.5\% immediately after training and 13\% after about 2 months, in the worst case. We also explore the use of our authentication approach for people with motor disabilities. Our analysis using a separate dataset of 6 subjects with non-spastic cerebral palsy shows an EER of 11.2\% immediately after training and 21.6\% after about 2 months, in the worst-case.
\end{abstract}

%%%%%%%%% BODY TEXT
\vspace{-0.1in}
\section{Introduction}
In recent years there has been an explosion in the proliferation of head-mounted wearable devices, with products such as smart eyewear (SEW) devices \cite{epson, glass,hololens,vuzix}, smart ear-buds \cite{misfit}, and smart headwear \cite{smartcap}. Most of these devices come with in-built accelerometer and gyroscope sensors, which can measure a variety of the wearer's head and body movements. One of the potential uses of such head-mounted wearable devices is to provide new sources of biometrics for authentication \cite{Wei*14}. %For example, these new biometrics can: (1) provide authentication for  Smart Eyewear devices (e.g., Microsoft Hololens), (2)  to their computing devices in addition to the traditional ones; and (3) provide people with motor disabilities to authenticate to their computing devices as they cannot make use of many traditional authentication solutions available to the rest of us such as passwords, fingerprints, face scan etc. 

Our principal {\bf goal} in this paper is to develop a passive authentication approach using head-mounted wearable devices (i.e., it works without requiring explicit action or gestures of any kind from the wearer). Passive authentication approaches are more difficult for adversaries to copy and spoof. Physiological signals (e.g., electroencephalogram (EEG)) make good biometrics for enabling passive authentication due to their inherently limited observability for adversaries. %However, typical head-mounted wearable devices do not come with physiological sensors. Hence, 
Our strategy is to use measurements from the two {\em movement sensors} (i.e., accelerometer and gyroscope) on head-mounted wearable devices 
% that are commonly available on such devices, 
and use them to derive something called a {\bf ballistocardiogram (BCG)}. BCG represents the body's motion as the blood flows through it, in response to the beating of the heart, and thus captures the characteristics of the cardiac cycle \cite{bioglass}. We derive the BCG used for authentication by asking each subject to sit still and then capturing and filtering the subtle and involuntary\footnote{Here, we use the term {\em involuntary} in the medical sense to mean {\em not under the conscious control of a person} \cite{md}.} head movements that their body makes as a result of the pumping action of the heart. In fact we divide the sensor measurements into fixed-duration {\em segments} and derive six versions of BCG (which we refer to as {\em BCG waveforms}), one for each axis of the accelerometer and gyroscope, from each segment. 

The BCG waveforms from several segments are then used to train a Convolutional Neural Network (CNN) classifier, which acts as the {\em authentication model}. Each authentication model is subject-specific (i.e., personalized), therefore different subjects have their own models. Once the models are trained, {\em authentication} is done by collecting one (or more) segment(s) of accelerometer and gyroscope measurements from an {\em unknown subject} wearing the head-mounted wearable device. The sensor measurements in the segment are filtered to derive six BCG waveforms, which are then fed into a subject-specific model (belonging to the subject in our system that the unknown subject claims to be). This model determines if these newly derived BCG waveforms are similar to the waveforms already seen from the subject earlier during model training. If deemed sufficiently similar, the unknown subject is then successfully authenticated.  We evaluate our authentication models using data from 12 subjects. Our approach demonstrated an equal error rate of 3.5\% immediately after training and 13\% after about 2 months, in the worst case. 

In addition, we also explore the use of our approach for subjects with motor disabilities. The idea was to see if it can provide an easy-to-use authentication alternative for a population that lacks full independence in using their computing devices securely. People with motor disabilities have tremendous difficulty existing authentication solutions (e.g., passwords, fingerprint), to the extent that they have to rely on others to authenticate for them \cite{Singh:2007:PSI:1240624.1240759}. Using a separate dataset of 6 subjects with non-spastic cerebral palsy, our approach demonstrated a worst-case equal error rate of 11.2\% after training and 21.6\% after about 2 months.

In this paper, we specifically use a Smart Eyewear (SEW) device (Google Glass \cite{glass}) as our head-mounted wearable device. We chose to use an SEW device because of its relative ubiquity and easy programmability compared to other head-mounted wearable devices. Note that our work is generalizable to any head-mounted wearable device and is in no way limited to SEW devices or to Google Glass. 

The {\bf contributions} of this paper are then three-fold. (a) A passive authentication approach that uses BCG derived from accelerometer and gyroscope measurements of subtle and involuntary head movements. (b) Demonstration of the viability of this authentication approach measurements collected from 12 able-bodied subjects over the course of approximately 3 months. (c) Demonstration of the promise of our authentication approach for people with motor disabilities, using measurements from 6 subjects with cerebral palsy over approximately 3 months.

\section{Problem Statement}
We next detail our problem statement and the assumed threat model for this work. The principal problem that we address in this paper is {\em to determine if ballistocardiogram derived from subtle head movements of a person wearing an SEW device (or more generally, a head-mounted wearable device) is capable of authenticating them.} %the current wearer of an SEW device (or more generally a head-mounted wearable device) is the person that they claim to be.}  
We assume that the {\bf threat} to our authentication approach comes from adversaries trying to declare themselves to be a particular subject (i.e., victim) and try to mimic their head movements to try to authenticate successfully. For the purposes of this work, we assume that adversaries: (1) do not have access to the authentication model, (2) cannot pollute the model during the training stage, and (3) do not have access to any form of cardiac signal from the victim's past or present.

%%%%%%%%%%%%%%%%%%%%%%%%%%%%%%%%%%%%%%%%%%%%%%%%%%%%%%%%%
\begin{figure*}
\centering
\includegraphics[width=1\textwidth,  trim={0 0 0 0}, clip]{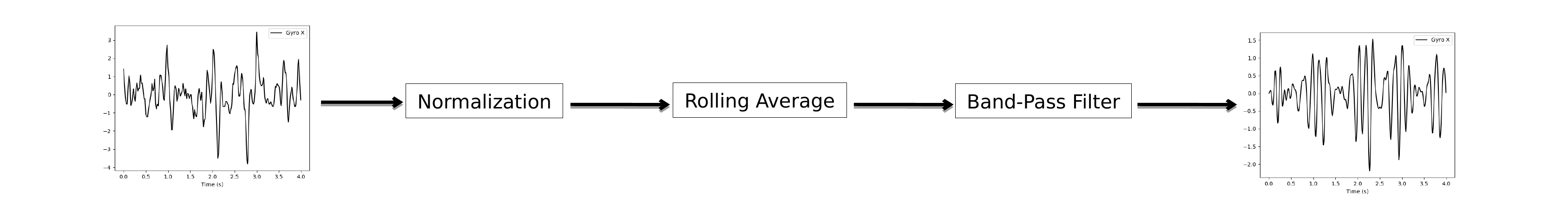}
\caption {{\bf The process of generating a BCG waveform from a segment of preprocessed x-axis gyroscope sensor stream. This process is performed on all three axes of accelerometer and gyroscope sensors.}}
\label{fig:bcg_gen}
\vspace{-0.1in}
\end{figure*}
%%%%%%%%%%%%%%%%%%%%%%%%%%%%%%%%%%%%%%%%%%%%%%%%%%%%%%%%

\section{Approach}
\label{sec:appr}
Our authentication approach has four stages. {\em Data collection and preprocessing} describes our process for gathering the accelerometer and gyroscope measurements from the SEW device followed by a preprocess step to synchronize them. {\em BCG extraction} shows the derivation of BCG from the head movement data. {\em Model training} uses the derived BCG to train a convolutional neural network (CNN) classifier as the authentication model. Finally, the {\em authentication process} itself uses the trained CNN classifier to authenticate the subject.

 % (1) {\em Data Collection:} During this stage we collect accelerometer and gyroscope measurements from the SEW device, thus capturing the involuntary subtle head movements of the wearer. (2) {\em BCG Extraction:} Here, we segment, clean, and filter  the movement sensor measurements to derive the BCG. (3) {\em Model Training and Authentication:} Finally, we train a convolutional neural network (CNN)-based authentication model based using the derived BCG, and then use the trained authentication model to identify the wearer of the SEW device at a future point in time. In the rest of this section, we describe these steps in detail.
 
\subsection{Data Collection Protocol} 
\label{sec:data_col}
The first stage in our authentication approach is to collect accelerometer and gyroscope measurements from subjects wearing the SEW device (a Google Glass in our case). We {\em standardize} the data collection protocol across the subjects to minimize motion artifacts and to ensure reproducibility of results. The data collection protocol used is as follows. We ask the subjects to sit comfortably, upright, and still. Their hands are placed on a table with their elbows forming a 90-degree angle. We then situate an SEW device on their faces such that the upper edge of the device aligns with their brow. We make sure that the device fits comfortably (not pinching or sitting unevenly) on the subjects' heads. Subjects wearing prescription glasses are asked to remove them to minimize SEW fitting issues. 

\subsection{Data Preprocessing} 
\label{sec:data_preproc}
During data collection, the accelerometer and gyroscope sensors in the SEW device are set to sample at 50 Hz. The SEW device relays measurements from the accelerometer/gyroscope sensors wirelessly to a nearby laptop, where the measurements are stored. As a result of the data collection process, we obtain six discrete, {\em raw sensor streams}: the three axes of the accelerometer and three axes of the gyroscope measurements. Sensor data from any Android device, like Google Glass, is not guaranteed to align exactly to a particular sampling rate. Therefore, a sampling rate of 50Hz resulted in an inter-sample-interval of anywhere from 5 to 20~ms. Additionally, there exists no guarantee that the samples recorded from the gyroscope and accelerometer measurements are synchronized or aligned in any way. In order to address these two concerns, we preprocess the sensor streams. In this regard, once the raw sensor streams are collected, we truncate the beginnings and endings of both the gyroscope and accelerometer measurements, such that the timestamp of the first and last samples of both sensor measurements are as close as possible. We then interpolate the data and align the samples with one another. The first sample of the gyroscope and the accelerometer now share the same timestamp, as do all the subsequent points in the sensor stream. All subsequent analyses use these {\em preprocessed sensor streams} instead of the raw sensor streams.

\subsection{Deriving the BCG Waveform} 
\label{sec:bcg_gen}
Once we obtain the preprocessed sensor streams of accelerometer and gyroscope measurements from individual subjects, the next step in our approach is to derive the BCG. To do this, we first divide each preprocessed sensor stream into  overlapped {\em segments} of size $w$ seconds. Between two sequential segments, there is a $w-1$ second overlap. Hence, two segments with $w=3$ seconds would share 2 seconds of data. Then, inspired by \cite{bioinsights}, we perform a three-step BCG derivation process. (1) {\em Normalization:} We normalize each of the six sensor streams to have a zero mean and unit variance within each segment. (2) {\em Rolling Average Filter:} We then subtract a rolling-average filter of 35 samples from each sensor stream to correct for large motions as well as gyroscope and accelerometer drift. (3) {\em Band-Pass Filter:} Finally, we apply a 4th-order band-pass Butterworth filter with cutoff frequencies at 4 and 11Hz to each sensor stream. Figure \ref{fig:bcg_gen} shows the main stages of BCG generation for one segment of a sensor stream. In all, we derive six versions of BCG (i.e., {\em BCG waveforms}), one per axis of accelerometer and gyroscope, {\em per segment}. These six BCG waveforms are then used as input for our authentication model.
%%%%%%%%%%%%%%%%%%%%%%%
\begin{figure*}
\centering
\includegraphics[width=1\textwidth,  trim={0 0 0 0}, clip]{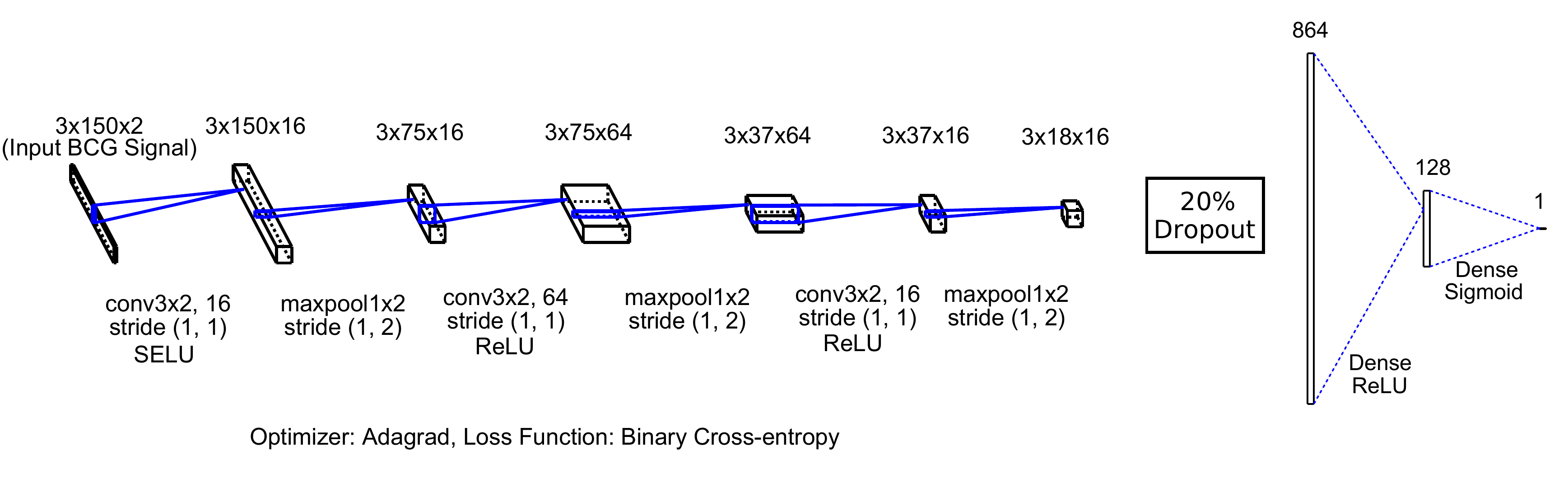}
\caption {{\bf Final CNN topology and the chosen parameters used in our subject-specific authentication model.}}
\label{fig:nn_topo}
\vspace{-0.1in}
\end{figure*}
%%%%%%%%%%%%%%%%%%%%%%%

\subsection{Model Training and Authentication}
We now construct an authentication model that learns the features of each subject's BCG waveforms and use their uniqueness to authenticate the subject at a later time. We use a convolutional neural network (CNN)-based classifier as our authentication model. We use CNNs due to their demonstrated ability to effectively classify time-series such as electrocardiogram (ECG), which, like BCG waveforms, are representations of the cardiac rhythm \cite{heartid}. In particular, our authentication approach has two stages, the training stage and the authentication stage.

{\em Training Stage:}  During the {\em training stage}, the goal is to enroll the subject into the authentication system by training a {\bf subject-specific authentication model} for them. This requires the collection of accelerometer/gyroscope measurements from a subject wearing the SEW device for $\Delta_{E}$ time-units. We use the 6 BCG waveforms from $\Delta/w$ segments (obtained from the subjects in our dataset (see Section \ref{sec:dataset})) to {\em train a CNN-based classifier} as an authentication model for a subject. During the training, we label the BCG waveforms from the subject's own segments as belonging to the positive class, while we consider the BCG waveforms from other subjects in out dataset to be part of the negative class. As such we use a one-versus-all strategy that allows each subject to get a custom, subject-specific model. 

In our CNN setup, the network has a particular emphasis on the relationships between the three axes of accelerometer- and gyroscope-derived BCG waveforms. To this end, we rearrange each $w$ segment to form a $2 \times 3 \times (w \times 50)$ tensor to be used as input to the CNN. Here, the first dimension refers to the measurement source (accelerometer or gyroscope), the second refers to the axis, and the third refers to time (where 50 is the sampling frequency in Hz). In designing our CNN we only fixed the inputs and outputs. As an output, the final layer is expressed as a single neuron reporting a value between 0.0 and 1.0, indicating the {\em confidence of the CNN} that the given sample should be accepted as belonging to the subject. In order to determine the intermediate layers of this CNN and their parameters, we take a {\em genetic algorithm-based approach}.  More details on the algorithm are given in Section \ref{sec:param_selection}. Figure \ref{fig:nn_topo} shows the final topology of the CNN we ultimately use. We train the CNN over the course of 100 epochs, at which point the loss of the model stabilizes to a minimum. At this point, the model has been trained and is ready to perform authentication.

{\em Authentication Stage:}\label{sec:auth} Once the model is trained, it is capable of performing {\em authentication}. To authenticate an unknown individual, we collect one (or more) $w$-second segment of raw accelerometer and gyroscope measurements while the unknown individual is wearing the SEW device. We derive BCG waveforms for each segment. These BCG waveforms are fed into the CNN of the authentication model, which produces a confidence score pertaining to how strongly it believes that the unknown individual is the same as the subject for whom the model was trained. If this confidence score is greater than a {\em decision threshold} $T$, the authentication is complete, and the subject is deemed to be authenticated. $T$ refers to the minimum confidence of the CNN required to accept a given input tensor.  A higher value of $T$ will make it more difficult to mistakenly accept negative-class BCG waveforms, but will also hinder the acceptance of positive-class BCG waveforms. 

The authentication step in our approach need not be a one-shot event; it can be repeated over several $w$-second segments. We use the variable $s$ to denote the number of segments (hence the number of {\em authentication attempts}) that were performed during authentication. In our approach, if any one of the $s$ sequential segments is accepted, the wearer is authenticated. Note that since between any two $w$-second segments there is a $w-1$-second overlap, a sequence of $s$ segments will only require $s+w-1$ seconds to measure.

\section{Experimental Setup}
\label{sec:es}
We now describe the dataset and metrics that we use to validate our approach, along with our experiments that establish the various parameters used in our approach.

%%%%%%%%%%%%%%%%%%%
%\begin{table}
%\small
%\centering
%\caption{\bf Data collection sessions and their spread w.r.t. the first session}
%\begin{tabular}{|c|c|} 
%\hline
%{\bf Session} & {\bf Time since 1st Data Collection Session} \\ 
% \hline
%Session 1 & 0  days \\  \hline
%Session  2  & 10-32 days \\  \hline
%Session 3 & 28-69 days \\ \hline
% \end{tabular}\\
% \vspace{-0.1in}
%\center{\footnotesize (The time-difference between any two sessions, for a subject, was at least 10 days.)}
%\label{tab:spread}
%\vspace{-0.2in}
%\end{table}
%%%%%%%%%%%%%%%%%%%
\subsection{Dataset}
\label{sec:dataset}
In order to validate our approach we collected data from 12 able-bodied volunteers. IRB approval was obtained for all data collection. From each subject we collected 10 minutes of accelerometer and gyroscope measurements using the Google Glass. 

Our goal with this project was to measure the {\em longitudinal effectiveness} of our authentication approach. Consequently, we collected {\bf three 10-minute sessions} from our subjects over approximately three months. During a session sitting, still for 10 minutes can be tedious; therefore, we broke up the 10-minute session into {\em five 2-minute intervals}. Between each interval subjects were given ample time to take a break and readjust themselves. During the 2-minute intervals, we asked the subjects to either focus on an eye-level sticky note stuck to the wall nearby, or to close their eyes. Further, we specifically asked the subjects not to focus on the screen of the Google Glass display as we found it is uncomfortable for anyone to do so for long periods of time. During any of the five 2-minute intervals, if the subjects moved in anyway, the data collection for that interval was stopped and repeated. 

All but one of the 12 subjects provided us with three sessions of data spread over 63 days. For every subject, the time-difference between any two sessions was at least 10 days. %Table \ref{tab:spread} shows, in days, the time  between the first data collection session and second and third data collection sessions.  
We refer to these 12 subjects as the {\em validation set}. In addition, we also collected data from 10 other subjects, which we call the {\em external set}. The data in the validation set are used to train the models for authentication. As we are building subject-specific models, we generate 12 models from the validation set. In contrast, the people in the external set include subjects from whom we only obtained one session (due to scheduling reasons) or those subjects whose data was collected in the pilot phase of the data collection. We use the external set for the evaluation of the ability of the subject-specific authentication models to reject data from subjects that it has never seen before. Table \ref{tab:demographics} summarizes the demographics of our dataset. 

%%%%%%%%%%%%%%%%%%%
\begin{table}
\small
\centering
\caption{\bf Dataset demographics}
\begin{tabular}{|c|c|c|c|c|} 
 \hline
 {\bf Set} & {\bf Avg. Age} & {\bf Std. Dev. Age} & {\bf \mars} & {\bf \female} \\ 
 \hline
 Validation & 32.83 & 13.13 & 4  & 8 \\ 
 External   & 28.50 & 10.91 & 7  & 3 \\
 \hdashline
 {\bf All}		& {\bf 30.86} & {\bf 12.09} & {\bf 12} & {\bf 10}\\ 
 \hline
\end{tabular}
\label{tab:demographics}
\vspace{-0.2in}
\end{table}
%%%%%%%%%%%%%%%%%%%

\subsection{Metrics}
In order to evaluate the efficacy of our approach we use the following core metrics: {\em false acceptance rate (FAR)}, {\em false rejection rate (FRR)}, and {\em equal error rate (EER)}.  FAR is the fraction of negatively labeled test BCG waveforms that were misclassified as positive. FRR is the fraction of positively labeled BCG waveforms that were misclassified as negative. Similarly, {\em true acceptance rate} (TAR) is the fraction of positively labeled BCG waveforms that were classified as positive, while {\em true rejection rate} (TRR) is the fraction of negatively labeled BCG waveforms that were classified as negative. The average of TAR and TRR provides us with {\em accuracy}. Finally, {\em equal error rate} (EER) is the rate at which FAR and FRR are equal. Even though we compute these metrics for every subject in our dataset, we present summary statistics of these metrics as an average over all subjects.  

\subsection{Parameter Selection}
\label{sec:param_selection} 
The data we use for all our parameter selection come from the first session of the subjects in the validation set. In order to tune our model parameters, we create 12 separate models, one for each of the subjects in the validation set. We then train the models using $\Delta_{E} = 8$ minutes of data, and tune using the remaining 2 minutes of data. Consequently, each model has 8 minutes of data from one subject in its positive class, and 88 minutes (8 minutes $\times$ 11 subjects) from other subjects in the validation set in the negative class. The remaining 24 minutes (2 minutes $\times$ 12 subjects) of data is used to test each model and tune its parameters based on the test results. All tests are done with $s=1$, i.e., with one-shot authentication using one segment of movement data. In our approach there are essentially two types of parameters that need to be decided: the parameters of the CNN and the segment length ($w$).

%%%%%%%%%%%%%%%%%%%
\begin{table}
\small
\centering
\caption{\bf Accuracy for different segment lengths ($w$)}
\begin{tabular}{|c|c|c|c|c|c|} 
 \hline
 {\bf segment length} & {\bf 1s} & {\bf 2s} & {\bf 3s} & {\bf 4s} & {\bf 5s} \\ 
 \hline
 Accuracy (\%) & 89.79 & 94.81 & {\bf 97.55} & 96.51 & 98.53\\ 
 \hline
\end{tabular}
\label{tab:w}
\vspace{-0.2in}
\end{table}
%%%%%%%%%%%%%%%%%%%
{\em Selecting the CNN's parameters:}  As mentioned before, we took a {\em genetic algorithm-based method} to find the parameters of our CNN-based authentication model. We optimized over a total of 10 traits that capture all the elements associated with convolutional and dense layers in the network. We omit the list of traits for clarity. Our algorithm first generates 20 CNNs, randomly selecting values from all possible traits and training them the first four 2-minute intervals (which forms $\Delta_{E}$) of the first session of data collection and evaluating their performance using the final 2 minutes of data in the first session. We use the scoring function $(FAR)^2+(FRR)^2$ to evaluate each CNN. In each generation we select the best 25\% CNNs as the parents for the next generation. Additionally, we select 3 CNNs at random from the bottom 75\% as parents. As such, the each generation has 8 parents. In order to replenish the next generation back up to 20, 12 children are created. This is done by selecting two of the parents at random from the pool of 8, and for each trait, randomly selecting between the two values held by the parents. A 15\% mutation rate is applied to the children in each generation. We ran this algorithm for 10 generations and selected the CNN with the overall best score as our model (shown in Figure \ref{fig:nn_topo}).

\begin{figure*}[t!]
    \centering
    \begin{subfigure}[b]{0.32\textwidth}
        \centering
        \includegraphics[width=.98\textwidth]{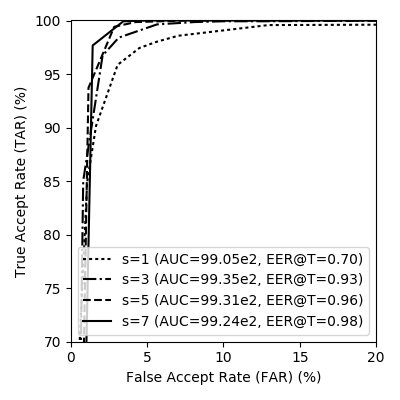}
        \caption{Session 1}
        \label{fig:roc1}
    \end{subfigure}
    \hfill
    \begin{subfigure}[b]{0.32\textwidth}
        \centering
        \includegraphics[width=.98\textwidth]{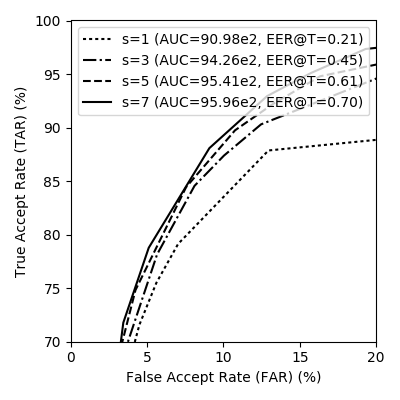}
        \caption{Session 2}
        \label{fig:roc2}
    \end{subfigure}
    \hfill
    \begin{subfigure}[b]{0.32\textwidth}
        \centering
        \includegraphics[width=.98\textwidth]{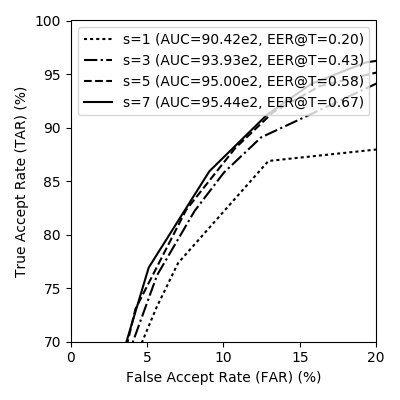}
        \caption{Session 3}
        \label{fig:roc3}
    \end{subfigure}
    \caption{\bf For each session, we generate ROC curves for each value of $s$ (authentication attempts) by varying $T$ (decision threshold). For visual clarity the x and y axis of the graphs have been shortened. The area-under-the-curve (AUC) values indicate the full extent of the performance of the models for various values of $s$.}   
    \label{fig:roc_all}
    \vspace{-0.2in}
\end{figure*}
%%%%%%%%%%%%%%%%%%%%%%%%

{\em Choosing the segment length $w$: } In order to determine the length of segments (i.e, $w$ seconds) of accelerometer and gyroscope measurements, from which we obtain the BCG waveforms, we evaluate values of $w$ from 1 second to 5 seconds in discrete steps using the data from the first session, as described previously. For each model, we compute the accuracy ((TAR+TRR)/2), and then average the accuracy across all models. We found that the the accuracy increases until $w=3$ seconds and then flattens, as shown in Table \ref{tab:w}. We therefore choose our segment length as $w=3$ seconds.
%%%%%%%%%%%%%%%%%%%%%%%%
\begin{figure}
\centering
\includegraphics[width=0.4\textwidth,  trim={0 13 0 0}, clip]{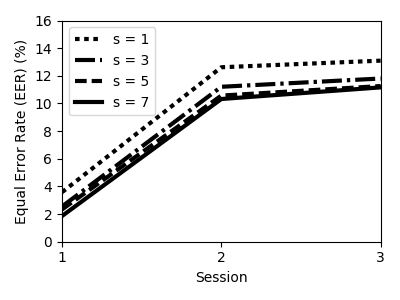}
\caption {\bf For each ROC curve, we isolate the EER. As expected, we see a degradation in EER as time between the training and authentication increases.}
\label{fig:eer-all}
\vspace{-0.2in}
\end{figure}
%%%%%%%%%%%%%%%%%%%%%%%%

\section{Evaluation}
\label{sec:t_and_s}
We now train a CNN classifier (parametrized with the the values chosen above) as the authentication model for each of the 12 subjects in our validation set. Subsequently, we {\em evaluate the efficacy of these models} in realistic settings, by using the yet-unseen samples of our dataset. This simulates the actions of the primary adversary of our threat model; someone who views a subject (i.e., victim) authenticating and then tries to mimic their head movements in order to authenticate successfully. As the victim does not make any head movement or gestures during authentication, the adversary has nothing to copy and is reduced to using their own head movements to try to authenticate.

\subsection{Evaluation Data Categories}
\label{sec:edc}
We evaluate our approach based on three categories of data from the dataset. (1) {\em Positive Validation:} Here, we test each subject's model with BCG waveforms derived from the subject's own yet-unseen (i.e., not used for training\footnote{This include segments from the final 2 minutes of data in session 1 along with segments from 10 minutes of data from sessions 2 and 3.}) segments. We use these positive-class BCG waveforms to compute the TAR (referred to as {\bf Validation TAR}) of our authentication approach. (2) {\em Negative Validation :} Similarly, we also test the BCG waveforms from rest of the segments in the validation set, i.e., those belonging to the other 11 users. The TRR (referred to as {\bf Validation TRR}), thus derived, demonstrates how well the model can prevent other subjects in the validation set from impersonating a particular subject. These BCG waveforms do not possess any temporal meaning with respect to the model being evaluated and are treated as one large set. (3) {\em Negative External:} Finally the external set is used to generate a second TRR (referred to as {\bf External TRR}), which demonstrates that our models have not overfit and are equally capable of denying entry to subjects whose data they have never seen in any form. Similar to (2), these BCG waveforms also do not possess any temporal meaning with respect to the model being evaluated.

Note that we have used the last 2 minutes of data from the subjects in the validation set to select the CNN parameters and choose the appropriate segment length. We maintain that this will not affect the correctness of our results because at no point are these last 2 minutes of data used in the actual training of models. %Furthermore, the inclusion of a far greater pool of longitudinal validation data should assuage any concerns of overfitting.

%%%%%%%%%%%%%%%%%%%%%%%%
\begin{figure*}[t!]
    \centering
    \begin{subfigure}[b]{0.42\textwidth}
        \centering
        \includegraphics[width=.8\textwidth]{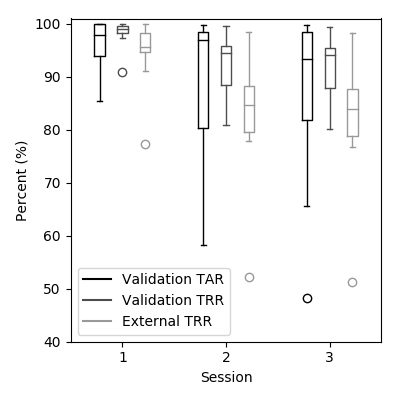}
        \caption{$s = 1$}
        \label{fig:box1}
    \end{subfigure}
    \hfill
    \begin{subfigure}[b]{0.42\textwidth}
        \centering
        \includegraphics[width=.8\textwidth]{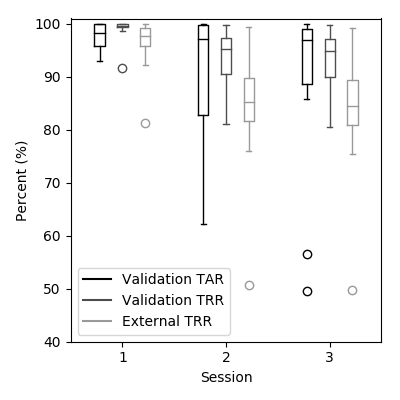}
        \caption{$s = 3$}
        \label{fig:box3}
    \end{subfigure}
    \hfill
    \begin{subfigure}[b]{0.42\textwidth}
        \centering
        \includegraphics[width=.8\textwidth]{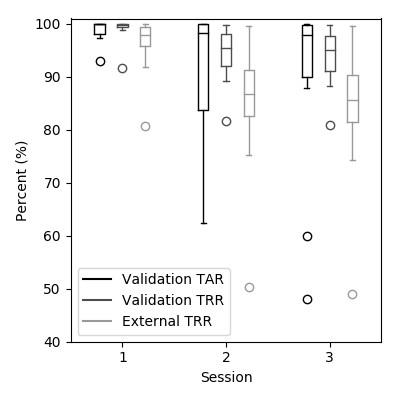}
        \caption{$s = 5$}
        \label{fig:box5}
    \end{subfigure}
     \hfill
    \begin{subfigure}[b]{0.42\textwidth}
        \centering
        \includegraphics[width=.8\textwidth]{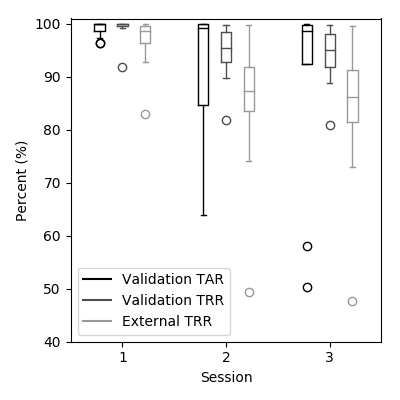}
        \caption{$s = 7$}
        \label{fig:box7}
    \end{subfigure}
    \caption{\bf Validation TAR, Validation TRR and External TRR for values of $s$  (authentication attempts) between 1 and 7, by using optimal values of $T$ over the three sessions. Note, the y-axis of the graph does not start at the origin.}
    \label{fig:boxen}
    \vspace{-0.2in}
\end{figure*}
%%%%%%%%%%%%%%%%%%%%%%%%

\subsection{Performance Analysis}
Figure \ref{fig:roc_all} shows the ROC curves for the 12 trained models, for multiple values of $s$, i.e., authentication attempts. The curves are produced by varying $T$, the decision threshold that provides the lower bound of the confidence level of the CNN. Here, $TAR$ = Validation TAR, and $FAR$ = $\{(1-$Validation TRR$)+(1-$External TRR$)\}/2$. We generate a total of three ROC graphs, one for each data collection session. It can be seen that, overall, our approach performs well. In session 1, the ROC curves show that the authentication models are very accurate with area-under-the-curve (AUC) values greater than $0.99$, irrespective of the number of authentication attempts. It is easy to see that the greater the number of authentication attempts, the higher the overall accuracy (i.e.,  $s=7$ always outperforms $s=1$). The accuracy of the authentication models drops in sessions 2 and 3. The performance drop between sessions 2 and 3 is, however, minimal as denoted by the AUC values. Since our segments are overlapped when $s=7$, it only requires $s+w-1$ seconds (i.e., 9 seconds) of sensor measurements.

The EER values for a given number of authentication attempts (i.e., value of $s$ ) is the point in Figure \ref{fig:roc_all}  where it's ROC curve meets the top left and bottom right diagonal (not shown in the figures) of the graph. Not surprisingly, the EER value occurs at different decision thresholds $T$. Figure \ref{fig:eer-all}  shows the EER values and its evolution over time. The magnitude of the EER value again shows the efficacy of our approach. For $s=1$, EER value is below 5\% in session 1, and then increases to around 13\% in sessions 2 and 3. Again higher values of $s$ produce lower EER values. %These results show that one of the future extensions of this approach could be to enable continuous authentication of wearer, something we plan to explore in the future.

Figure \ref{fig:boxen} shows how the Validation TAR, Validation TRR, and External TRR vary for our 12 subjects over the three sessions when the models are set to the decision threshold $T$ where we achieve EER.  It can be seen that for session 1, the box plots are very tight for all $s$. However, as we move to sessions 2 and 3, we vary the threshold $T$ such that we remain at the EER. This ensures that our TAR does not drop too precipitously, while still keeping a reasonably high TRR. The External TRR has a higher spread than Validation TRR, however, the medians of both are still very similar to each other. Again, as the number of authentication attempts ($s$) increase, the accuracy increases overall as seen from the lower spread of the box plots. %We can see that by updating $T$ over time, we can optimize our solution for improved longitudinal accuracy. %However, determining what $T$ value to use at any given time, in a general sense, will require a much larger and representative dataset.

These results show the viability of using BCG, collected using head-mounted wearable devices, for authentication.

%This being said, however, it is not our intention to present the selections of $T$ and $s$ shown here to be optimal in all scenarios. Rather, it is to show the process for optimizing our solution for temporally longitudinal application. More generally, we have shown that with a larger sample population, it is trivial to then tune these external parameters for optimal performance, such that our solution becomes generalizable.

\subsection{Performance for Individuals with Motor Disabilities}
We also deployed our authentication approach on a dataset where the subjects had severe motor disabilities. Such individuals cannot independently authenticate to their computing devices because traditional authentication approaches, like passwords, were not designed for such populations, and therefore, are extremely difficult to use \cite{Singh:2007:PSI:1240624.1240759}. The passive nature of our approach has the potential to allow individuals with motor disabilities to authenticate to their devices with minimal effort and therefore increase their independence in using modern computing devices.
%%%%%%%%%%%%%%%%%%%%%%%%
\begin{figure}
\centering
\includegraphics[width=0.4\textwidth,  trim={0 12 0 0}, clip]{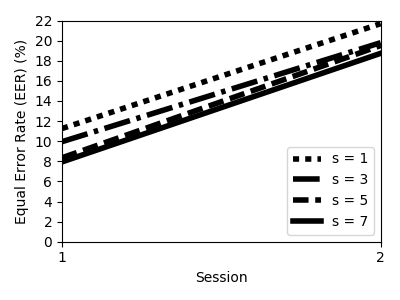}
\caption {\bf EER calculated from models trained with data from subjects with motor disabilities.}
\label{fig:eer-d}
\vspace{-0.3in}
\end{figure}
%%%%%%%%%%%%%%%%%%%%%%%%

We collected a separate dataset with 6 subjects (average age of 40 (11.06 stdev), 3 female, 3 male) with non-spastic cerebral palsy for this portion of the work. We obtained the requisite IRB approval and collaborated with a local non-profit organization to obtain the data. Due to scheduling difficulties, we collected {\em two sessions} of data from every subject but one. Further, the second session was conducted after 15 days for two subjects and 57 days for the other three, after the initial session.

Once again, we created subject-specific authentication models and repeated the evaluation process described in Section \ref{sec:t_and_s} for each of these six subjects. All the parameters for our authentication model were identical to those described in Section \ref{sec:es}. We evaluated these models using the external set of completely unseen data mentioned in Section \ref{sec:dataset}, along with Validation Positive and Validation Negative categories, as described in Section \ref{sec:edc}, created using the data from these six subjects.

Figure \ref{fig:eer-d} shows the EER results for this dataset over the two sessions. Overall, we observe a degradation in performance compared to the larger dataset. The worst case EER is 11.2\% immediately after training and 21.6\% after around 2 months, again both for $s=1$. We attribute this to a number of factors: (1) we observed that this group had increased difficulty in maintaining a motionless posture, given that their condition often affects their head muscles as well, (2) many of them were assisted by medical devices that would, on occasion, generate vibrational noise, and (3) we did not tune the CNN for this particular population. That being said, note that for $s=7$ the EER values were at least 3-4\% lower and we suspect further increasing $s$ and tuning the CNN would lower the EER even more. Given the poor state of authentication solutions for people with motor disability we view these results as a promising first step.

%One note on this approach; in this scenario, we only used the subjects with disabilities as the training pool, as opposed to using them in addition to our larger pool. The rationale for this was largely that we wanted to avoid the model learning to discriminate based on the random noise generated by this group, as opposed to the largely still main group. In this way, we demonstrate that the model is able to effectively differentiate subjects even when their physiological conditions are similar.

\section{Discussion}
The results demonstrate the potential of our approach. However, this work has some limitations, which plan to address in the future. These include:

{\bf Compensating for various types of artifacts.} Motion artifacts are a big problem with extracting physiological responses from movement data \cite{bioglass}. Therefore, we need strategies to compensate for noise induced in the signal from the motion of the user. Particularly: (1) when used in locations with heavy foot traffic and (2) when used while the user performs different activities (e.g., standing, walking etc.). 

{\bf Potential implementation issues.} This paper only describes an analysis of our authentication approach, not its implementation. Ideally, we want the authentication decision to be made locally on the head-mounted wearable device, to minimize the security risks of off-loading the authentication decision. Implementing our approach on head-mounted wearable devices needs to be explored further.

{\bf Effects of fatigue and physical activity:} In our current dataset the subjects were alert during all data collection sessions. However, it has been shown that factors such as fatigue or recent physical activity, effect an individual's physiology, movements, and posture \cite{stern}. At this time it is not clear how well our approach works when authenticating individuals who are fatigued, under the weather, or have engaged in recent physical activity. 

{\bf Model re-training schedule.} As physiological responses from subjects change over time, we see that our authentication accuracy drops in sessions 2 and 3.  Adjusting the decision threshold will help to reduce authentication errors only to some extent. At some point we have to retrain the models to capture the current physiology of the individual. Approaches are required to determine when to retrain the authentication models, such that the overall drop in authentication accuracy of the models is balanced with the inconvenience of taking the system offline.

\section{Related Work}
Ballistocardiography has been tried for authentication purposes \cite{Guo2013, bioinsights, 6617172}. In \cite{Guo2013, 6617172}, ballistocardiography was used on movement data collected from the person's torso, which makes it substantially easier to accurately detect the person's cardiac properties. The nature of ballistocardiography precludes us from using these results because as we move away from the heart, the noise in the derived BCG waveform increases dramatically \cite{bioinsights}. This is borne out by the fact that the original attempt to identify a subject based on BCG signals measured using hand-worn movement sensors, produced only a 66\% accuracy rate \cite{bioinsights}. 

Authentication approaches have been previously explored in the context of SEW devices as well. In \cite{Percom}, the authors use the notion of head movement in response to a specific song as a signature for authentication. The head movements used by this system are, however, very simple and can be easily spoofed by an adversary. In \cite{Schneegass*16}, the authors induce white noise into the subject's skull, the response to which is then picked up to determine who is wearing the device. This is a better solution in terms of spoofing resistance, but it requires the use of bone conductance speakers, which are something that not all SEW devices possess. This approach requires around 23 seconds of inducing noise before it can identify the subject, which is too slow to be practical. Further, the use of white-noise for authentication has been found to be uncomfortable to some subjects, as noted by the authors. In \cite{Rogers*15}, the authors present an approach for user identification in SEW devices, which uses the blinking and head movement pattern of the subject while they watch a short video on the device's screen. However this approach requires 34 seconds to identify the user, presenting an obvious temporal barrier to usefulness.

Furthermore, none of the existing work using ballistocardiography for authentication or SEW-device authentication has tested their models longitudinally, something we explicitly address in this work.

\section{Conclusions}
In this paper we have demonstrated a new authentication approach using ballistocardiogram (BCG) collected from head-worn wearables. We use BCG waveforms derived from accelerometer and gyroscope sensor measurements in a Smart Eyewear device, and achieved an EER of $\sim$4\% immediately after training, and $\sim$13\% after almost 60 days of training.  Additionally, we demonstrated that our approach holds promise as a new authentication option for individuals with motor disabilities. % to increase their level of independence in securely using their computing devices. 

In the {\em immediate future} we plan to extend this work in several directions including: (1) studying the use of BCG-based authentication for a larger population of people with motor disabilities, (2) making the approach tolerant to the presence environmental noise (e.g., foot traffic), (3) evaluating the consequences of recent physical activity on our authentication accuracy, and (4) implementing our approach on the SEW device.

{\small
\bibliographystyle{ieee}
\bibliography{biblio,bib4,hmd,iswc,cns,egbib}
}

\end{document}